\documentclass[sigconf]{acmart}

\setcopyright{none}
\acmConference[WSDM '27]{The 20th ACM International Conference on Web Search and Data Mining}{February 15--19, 2027}{Hong Kong}
\acmYear{2027}
\renewcommand\footnotetextcopyrightpermission[1]{}

\usepackage{booktabs}
\usepackage{array}
\usepackage{microtype}
\usepackage{xcolor}
\usepackage{graphicx}
\usepackage{amsmath}
\usepackage{multirow}

\author{Jinfeng Xu}
\email{jinfeng@connect.hku.hk}
\authornote{Corresponding author.}
\affiliation{%
  \institution{The University of Hong Kong}
  \city{HongKong SAR}
  \country{China}}

\author{Zheyu Chen}
\email{zyueee.chen@connect.polyu.hk}
\affiliation{%
  \institution{The Hong Kong Polytechnic University}
  \city{HongKong SAR}
  \country{China}}

\author{Ziyue Peng}
\email{zpengbe@connect.ust.hk}
\affiliation{%
  \institution{The Hong Kong University of Science and Technology}
  \city{HongKong SAR}
  \country{China}}

\author{Zheng Lin}
\email{zhenglin@ieee.org}
\affiliation{%
  \institution{University of Luxembourg}
  \country{Luxembourg}}

\author{Wenhao Yuan}
\email{wenhao.yuan@connect.hku.hk}
\affiliation{%
  \institution{The University of Hong Kong}
  \city{HongKong SAR}
  \country{China}}

\author{Jian Chen}
\email{ccccccj03@connect.hku.hk}
\affiliation{%
  \institution{The University of Hong Kong}
  \city{HongKong SAR}
  \country{China}}

\author{Shujie Li}
\email{shujie.li@connect.hku.hk}
\affiliation{%
  \institution{The University of Hong Kong}
  \city{HongKong SAR}
  \country{China}}

\author{Edith C. H. Ngai}
\authornotemark[1]
\email{chngai@hku.hk}
\affiliation{%
  \institution{The University of Hong Kong}
  \city{HongKong SAR}
  \country{China}}

\newcommand{\cga}{\ensuremath{\mathrm{CGA}}}
\title{Beyond Successor Accuracy: State Retention for Recursive Self-Improvement in Recommendation}

\begin{document}

\begin{abstract}
Recommendation recursive self-improvement (Rec-RSI) feeds recommender outputs into subsequent training. Evaluating each round solely through its latest model assumes that the successor consolidates the update, although pre- and post-update models may retain complementary ranking decisions. We term this \emph{distributed progress} and quantify it using cross-generation advantage (CGA), a marginally matched contrast between cross- and within-generation model pairs. A rank-separation statistic, label-free at selection time, predicts which family to retain. Across four datasets and three sequential recommendation encoders, the preferred retention regime varies by architecture: cross-generation pairing benefits GRU4Rec and SASRec, whereas FMLP initially favors within-generation pairing and shifts toward cross-generation pairing after a second update. Rank separation selects the stronger family in 12/12 first-update and 5/6 second-update dataset--encoder settings; on held-out tests, the selected family outperforms the direct successor in 34/36 trajectories. Five transfer mechanisms do not consistently reproduce these gains in one model. These findings establish state retention as a distinct Rec-RSI problem: progress may reside in relations between generations as well as in the latest model. Code is available at \href{https://github.com/Jinfeng-Xu/RecRSI}{https://github.com/Jinfeng-Xu/RecRSI}.
\end{abstract}

% \ccsdesc[300]{Computing methodologies~Machine learning}

\maketitle

\section{Introduction}
Recommendation recursive self-improvement (Rec-RSI) places a recommender in a learning loop that generates supervision, trains a successor, and lets that successor shape later updates. Recent systems instantiate this loop through fidelity-controlled interaction generation~\cite{RSIR} or agent-based revision of objectives, code, and pipelines~\cite{SelfEvolveRec,YouTubeSelfEvolving}. Each update changes both the ranking policy and the evidence available to future generations. Evaluating each round only through the latest model adopts a replacement view, attributing progress to the successor whenever Recall or NDCG rises. Yet these aggregate metrics obscure which relevant items support the score. Models with similar utility may retrieve different items, and an improved successor may not preserve all useful predecessor behavior. Since the retained state governs the next round, Rec-RSI raises a question beyond successor accuracy. What should persist across an update?

We posit that recursive progress can be relational. An update may create complementarity across its boundary even when no individual model contains the joint behavior. We call this stage-structured complementarity \emph{distributed progress} and ask whether cross-generation pairs provide greater utility than pairs formed within either generation.

\begin{figure*}[t]
  \centering
  \includegraphics[width=\textwidth]{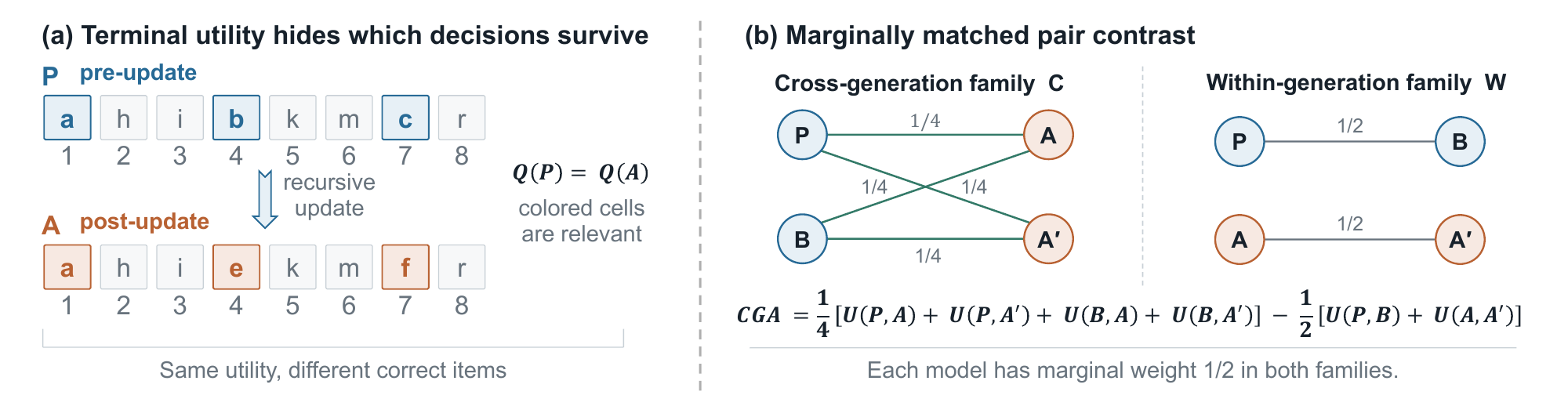}
   \vskip -0.2in
  \Description{Panel a shows a pre-update and post-update recommender with similar aggregate utility but different correct top-ranked items. Panel b retains models from both generations and compares cross-generation fusion with within-generation fusion under matched model marginals.}
  \caption{Successor accuracy can obscure where useful behavior persists. (a) Similar aggregate utility may arise from different relevant top-ranked items. (b) CGA contrasts cross-generation pairs, including parent--child and non-genealogical pairs, against marginally matched within-generation pairs.}
  \label{fig:measurement}
  \vskip -0.15in
\end{figure*}

We quantify this state with cross-generation advantage (CGA), which contrasts cross- and within-generation pairs under matched marginal exposure and equal inference cost. The contrast targets the interaction associated with crossing the update boundary. Rank separation then uses earlier-window top-$k$ overlap to select a pair family without outcome labels at selection time. Figure~\ref{fig:measurement} illustrates the relational state and its matched contrast. Across four datasets, three encoders, and two recursive updates, retention regimes vary by architecture. Rank separation predicts these regimes, and the selected family outperforms the direct successor in 34/36 held-out trajectories. Five transfer routes spanning data, predictions, rankings, targets, and parameters fail to consolidate these gains reliably, identifying relational state retention as a central Rec-RSI problem.

\section{Relational State across Updates}

\subsection{Single-Model Consolidation}

In Recommendation RSI, progress may be encoded in generated supervision, predictive distributions, ranked lists, complementary targets, or parameters. To study whether a single successor can consolidate such progress, let $z$ denote one source configuration, $z_0$ its matched control, and $Q(z)$ the ranking utility obtained in its native form. The source gain is $\Delta_{\mathrm{src}}=Q(z)-Q(z_0)$. A transfer operator $T$ consolidates each source into one successor, whose gain is $\Delta_{\mathrm{succ}}=Q(T(z))-Q(T(z_0))$. The residual: $\Delta_{\mathrm{src}}-\Delta_{\mathrm{succ}}$, quantifies the gain left outside that successor. We instantiate $T$ with five common consolidation routes.

\subsection{Cross-Generation Advantage}

If transfer is incomplete, useful behavior may remain across update stages. Each trajectory contains four models. Models $P$ and $B$ are independent pre-update runs. Model $P$ generates the interactions used to train post-update model $A$; $A'$ is trained on the same data with a different seed. For item $i$ and user $u$, we fuse two full-catalog distributions as $S_\beta(i\mid u;X,Y)=(1-\beta)\log p_X(i\mid u)+\beta\log p_Y(i\mid u)$ and rank by $S_\beta$, with $\beta=0.5$. 
Let $U(X,Y)$ be the resulting ranking utility. The cross- and within-generation families are:
\begin{align*}
\mathcal{C}&=\{(P,A),(P,A'),(B,A),(B,A')\},\\
\mathcal{W}&=\{(P,B),(A,A')\}.
\end{align*}

We define:
\begin{equation}
 \cga=\frac{1}{4}\sum_{(X,Y)\in\mathcal C}U(X,Y)
 -\frac{1}{2}\sum_{(X,Y)\in\mathcal W}U(X,Y).
 \label{eq:cga}
\end{equation}
Each side assigns total weight $0.5$ to $P,A,B,$ and $A'$. Under an additive decomposition of pair utility into constituent main effects and an interaction, matching the marginals cancels the main effects. CGA then captures the difference between average cross- and within-generation interaction terms. Under the fixed fusion rule, this interaction can reflect calibration, score scale, and top-rank decisions without being assigned to one mechanism. Positive \cga{} means that cross-generation pairing yields higher utility. Because $\mathcal C$ contains both genealogical and non-genealogical pairs, \cga{} characterizes the update as a whole rather than a specific parent--child edge.

\subsection{Rank Separation for State Selection}

For model $X$, let $L_X^k(u)$ be user $u$'s top-$k$ list, and define $O_k(X,Y)=\mathbb{E}_u[|L_X^k(u)\cap L_Y^k(u)|/k]$. We measure whether the update separates rankings more than seed variation:
\begin{equation}
 \mathrm{Sep}_k=\frac{1}{|\mathcal W|}\sum_{(X,Y)\in\mathcal W}O_k(X,Y)
 -\frac{1}{|\mathcal C|}\sum_{(X,Y)\in\mathcal C}O_k(X,Y).
 \label{eq:separation}
\end{equation}

Positive $\mathrm{Sep}_{20}$ means that adjacent generations disagree near the recommendation boundary more than do two models from the same generation. Computing it at selection time requires no relevance labels.

\section{Evaluation Protocol}

\paragraph{Datasets and encoders.}
We use Amazon Toys, Amazon Beauty, Yelp, and Amazon Sports with GRU4Rec~\cite{GRU4Rec}, SASRec~\cite{SASRec}, and FMLP~\cite{FMLPRec}, spanning recurrent, attention, and frequency-domain sequence encoders. The evaluation cohorts contain 9,011, 10,980, 17,168, and 17,297 users, respectively. Each dataset--encoder setting has three independently seeded trajectories. 

\paragraph{First-update protocol.}
Each trajectory trains pre-update models $P,B$ and post-update models $A,A'$ for 10,000 updates. Model $P$ generates the interactions used by $A$ and $A'$. Models $A$ and $A'$ share the same data and configuration but use different seeds. Here $m=5$ produces five continuations per fit history, $k=20$ is the upstream rank-fidelity threshold, and pair fusion fixes $\beta=0.5$. The final three events define windows for checkpoint selection, diagnosis, and confirmation; earlier events form the fit prefix, and only the selection window chooses checkpoints. Same-data seeds control for initialization. The snapshot control is the best of four within-model pairs: 5k--10k and best--final checkpoints for $P$ and $A$. Each uses the same two-model computation and fusion as the candidate families. The protocol and indicator were specified on Toys, Beauty, and Yelp; Sports was evaluated prospectively, while the original validation and test splits were used only for final evaluation.

\paragraph{Second-update extension.}
On Toys, Beauty, and Sports, model $A$ generates $D_2$ from the fit prefix augmented with $D_1$. Two new seeds, $C$ and $C'$, then train on the fit prefix augmented with $D_1$ and $D_2$ using the same 10,000-update budget. We denote the post-update generations after $D_1$ and $D_2$ by G1 and G2, and compare their adjacent families on train-derived windows. This yields 18 second-update trajectories across GRU4Rec and FMLP.

\begin{table}[t]
\centering
\caption{Original held-out test (three-trajectory mean). W and C denote within- and cross-generation families. CGA is NDCG@20 ($\times10^{-3}$), followed by positive trajectories. Bold follows the unrounded values.}
\vskip -0.15in
\label{tab:main}
\scriptsize
\resizebox{\linewidth}{!}{%
\begin{tabular}{@{}cccccccccc@{}}
\toprule
\multirow{2.5}{*}{Encoder}
& \multirow{2.5}{*}{Metric}
& \multicolumn{2}{c}{Toys}
& \multicolumn{2}{c}{Beauty}
& \multicolumn{2}{c}{Yelp}
& \multicolumn{2}{c}{Sports} \\
\cmidrule(lr){3-4}
\cmidrule(lr){5-6}
\cmidrule(lr){7-8}
\cmidrule(l){9-10}
& & W & C & W & C & W & C & W & C \\
\midrule

\multirow{4.5}{*}{GRU4Rec}
& R@10
& .0203 & \textbf{.0210}
& .0446 & \textbf{.0458}
& .0271 & \textbf{.0281}
& .0196 & \textbf{.0209} \\
& N@10
& .0101 & \textbf{.0105}
& .0240 & \textbf{.0247}
& .0137 & \textbf{.0140}
& .0098 & \textbf{.0106} \\
& R@20
& .0344 & \textbf{.0354}
& .0683 & \textbf{.0700}
& .0460 & \textbf{.0477}
& .0326 & \textbf{.0347} \\
& N@20
& .0136 & \textbf{.0141}
& .0300 & \textbf{.0308}
& .0184 & \textbf{.0189}
& .0131 & \textbf{.0141} \\
& CGA
& \multicolumn{2}{c}{+.442 (3/3)}
& \multicolumn{2}{c}{+.789 (3/3)}
& \multicolumn{2}{c}{+.466 (3/3)}
& \multicolumn{2}{c}{+.951 (3/3)} \\
\midrule
\addlinespace[0.5pt]

\multirow{4.5}{*}{SASRec}
& R@10
& .0532 & \textbf{.0533}
& .0667 & \textbf{.0676}
& .0319 & \textbf{.0324}
& .0313 & \textbf{.0319} \\
& N@10
& .0302 & \textbf{.0304}
& .0383 & \textbf{.0387}
& .0162 & \textbf{.0164}
& .0177 & \textbf{.0179} \\
& R@20
& .0748 & \textbf{.0752}
& .0923 & \textbf{.0932}
& .0533 & \textbf{.0539}
& .0465 & \textbf{.0474} \\
& N@20
& .0357 & \textbf{.0359}
& .0448 & \textbf{.0451}
& .0216 & \textbf{.0218}
& .0215 & \textbf{.0218} \\
& CGA
& \multicolumn{2}{c}{+.280 (3/3)}
& \multicolumn{2}{c}{+.342 (3/3)}
& \multicolumn{2}{c}{+.238 (3/3)}
& \multicolumn{2}{c}{+.277 (3/3)} \\
\midrule
\addlinespace[0.5pt]

\multirow{4.5}{*}{FMLP}
& R@10
& \textbf{.0466} & .0466
& .0625 & \textbf{.0625}
& .0224 & \textbf{.0226}
& .0235 & \textbf{.0235} \\
& N@10
& .0275 & \textbf{.0276}
& .0381 & \textbf{.0381}
& .0115 & \textbf{.0115}
& \textbf{.0132} & .0132 \\
& R@20
& \textbf{.0618} & .0612
& \textbf{.0838} & .0834
& \textbf{.0371} & .0368
& \textbf{.0341} & .0335 \\
& N@20
& \textbf{.0314} & .0313
& \textbf{.0435} & .0434
& \textbf{.0151} & .0150
& \textbf{.0159} & .0157 \\
& CGA
& \multicolumn{2}{c}{-.063 (2/3)}
& \multicolumn{2}{c}{-.044 (1/3)}
& \multicolumn{2}{c}{-.098 (1/3)}
& \multicolumn{2}{c}{-.240 (0/3)} \\
\bottomrule
\end{tabular}%
}
\vskip -0.2in
\end{table}

\paragraph{Consolidation baselines and training.}
The consolidation study evaluates hard generated-data retraining, full-catalog KL distillation, top-50 listwise distillation, complementary item targets, and linear parameter interpolation. Hard retraining is compared with retraining from independent teacher $B$; both distillation routes with an $A$-only student; complementary targets with context-matched random targets; and interpolation with the stronger endpoint. Shared pairs are averaged first within each trajectory; dataset--encoder settings are the primary comparison units, while trajectory rates are descriptive. Effect intervals are computed from 20,000 dataset-cluster bootstrap replicates. All models use 64-dimensional item embeddings. GRU4Rec has two recurrent layers (hidden size 256, dropout 0.2); SASRec has two attention layers (two heads, hidden size 128, dropout 0.5); FMLP has two filter layers (dropout 0.5). We use Adam at $10^{-3}$, a batch size of 256, 256 sampled negatives, and a maximum sequence length of 50; evaluation runs every 500 updates. Full-catalog evaluation masks padding and observed items. Users must have at least four original training events.

\section{Empirical Results}

\subsection{Distributed Progress after Consolidation}

The direct post-update model $A$ exceeds its parent $P$ in 26/36 trajectories on the original held-out test. Improved successor accuracy therefore does not preclude distributed progress, as the pairwise results show. The five consolidation routes recover different portions of that progress: generated-data retraining changes sign across domains; full-catalog, listwise, and complementary-target transfer beat their controls in 1/4, 2/4, and 1/4 settings, respectively; parameter interpolation never beats its stronger endpoint in ten trajectories. Listwise transfer retains the most, reducing the mean source-to-successor loss from $2.313\times10^{-3}$ to $1.375\times10^{-3}$ NDCG@20, while leaving a measurable share of the source gain distributed across models.

\begin{figure}[t]
  \centering
  \includegraphics[width=1\columnwidth]{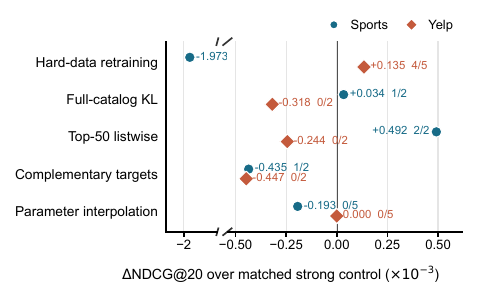}
  \vskip -0.2in
  \Description{A broken-axis dot plot compares five routes for consolidating recursive state into one model on Sports and Yelp. Every route has a negative result in at least one domain, and most change sign across domains.}
  \caption{Single-model consolidation varies across domains. Points show $\Delta$NDCG@20 relative to route-specific comparators ($\times10^{-3}$); labels report the value and wins among applicable trajectories.}
  \vskip -0.2in
  \label{fig:compression}
\end{figure}

\subsection{Encoder-Dependent Pair Utility}

Table~\ref{tab:main} reports the matched families. The cross-generation family improves GRU4Rec and SASRec on every dataset at both cutoffs and under both metrics; all eight NDCG@20 setting means are positive with 3/3 trajectory wins. FMLP reverses this ordering at $k=20$, while its $k=10$ differences remain near zero. The NDCG@20 family preference agrees with the confirmation window in all 12 settings. Separately, FMLP's cross-generation family exceeds the snapshot control in 11/12 test trajectories, while its negative CGA favors the matched within-generation family.

\subsection{Rank Separation as a Retention Signal}

\begin{figure}[t]
  \centering
  \includegraphics[width=\columnwidth]{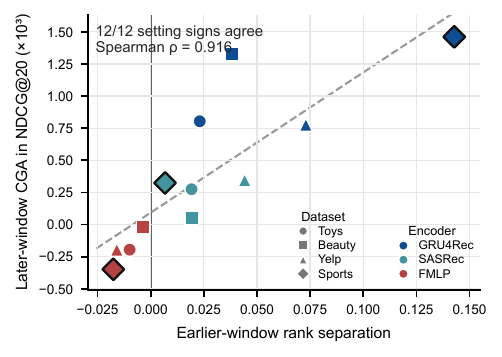}
  \vskip -0.2in
  \Description{A scatter plot shows that rank separation measured on an earlier window marks later cross-generation advantage. GRU4Rec and SASRec occupy the positive region, while FMLP occupies the negative region across four datasets; Sports points are outlined.}
  \caption{On the first update, rank separation marks the \cga{} sign on nine development settings and all three Sports settings (outlined).}
  \vskip -0.2in
  \label{fig:stage}
\end{figure}

Figure~\ref{fig:stage} connects earlier-window rank separation to later-window \cga{}. The zero-threshold indicator retains the cross-generation family when $\mathrm{Sep}_{20}>0$ and the within-generation family otherwise. $\mathrm{Sep}_{20}$ and CGA have the same sign in all nine development settings and all three Sports settings; across the 12 settings, Spearman $\rho=0.916$. Leave-one-dataset-out and leave-one-encoder-out analyses preserve every remaining setting sign ($9/9$ and $8/8$, respectively), with corresponding $\rho$ ranges of $0.850$--$0.930$ and $0.710$--$0.930$. On the development datasets, the correlations of \cga{} with top-20 separation and unique-hit advantage are $0.850$ and $0.870$, respectively, compared with $0.500$ and $0.270$ for full-catalog Jensen--Shannon and total-variation separation.

This contrast locates the predictive signal near the ranking boundary: entry into the exposed list is more informative about later pair utility than global agreement over the full catalog. The association is primarily between architectures rather than between seeds within an architecture. Across individual trajectories, the correlation is $0.814$ but falls to $0.205$ after ranks are centered within each architecture. Rank separation therefore diagnoses how an encoder responds to the current update; it is not a reliable selector among seeds of the same encoder.

\subsection{Generalization across Datasets and Update Rounds}

At the setting level, the indicator chooses the stronger family in all 12 first-update settings, including 3/3 Sports settings; at the trajectory level, it succeeds in 32/36 cases. The selected family beats the snapshot control in 35/36 trajectories by $2.065\times10^{-3}$ NDCG@20, with a 95\% dataset-cluster bootstrap interval of $[1.597, 2.512]\times10^{-3}$.

\begin{table}[t]
\centering
\caption{Selection-time label-free decisions. Setting and Traj. count correct family choices; Snap. counts wins over the snapshot control. Gain is selected-family minus snapshot-control NDCG@20 ($\times10^{-3}$).}
\vskip -0.15in
\label{tab:gate}
\small
% \footnotesize
% \setlength{\tabcolsep}{3.0pt}
% \renewcommand{\arraystretch}{1.04}
\begin{tabular}{@{}lrrrr@{}}
\toprule
Evidence & Setting & Traj. & Snap. & Gain \\
\midrule
Development & 9/9 & 24/27 & 26/27 & +1.884 \\
Sports (prospective) & 3/3 & 8/9 & 9/9 & +2.605 \\
All first update & 12/12 & 32/36 & 35/36 & +2.065 \\
Second update & 5/6 & 15/18 & 17/18 & +2.057 \\
\bottomrule
\end{tabular}
\vskip -0.15in
\end{table}

The original held-out test preserves the architecture sign in 12/12 settings, and setting effects track the confirmation window ($\rho=0.921$, Pearson $r=0.963$). The selected family exceeds successor $A$ in 34/36 trajectories by $2.018\times10^{-3}$ NDCG@20 on average (a 10.6\% relative gain), with a dataset-cluster interval of $[1.380, 2.613]\times10^{-3}$. It also exceeds the best single model in 33/36 trajectories by $1.457\times10^{-3}$; retaining the pair adds one stored model and one forward pass.

The second update is not a replay of the first. GRU4Rec preserves positive CGA on all three datasets and in all 9 trajectories, although its pooled magnitude falls from $1.196\times10^{-3}$ to $0.749\times10^{-3}$. FMLP's pooled CGA moves from $-0.187\times10^{-3}$ to $+0.035\times10^{-3}$, while $\mathrm{Sep}_{20}$ becomes positive in all 18 trajectories. The indicator selects 5/6 setting-level and 15/18 trajectory-level regimes, and beats the G2 snapshot control in 17/18 trajectories by $2.057\times10^{-3}$. G2 thus shifts toward cross-generation divergence, while realized CGA remains setting-dependent (5/6 settings).

% \paragraph{Implications for Recommendation RSI.}

% Recommendation RSI commonly carries one artifact forward: a dataset, objective, program, or successor. CGA reveals that the operative state may instead be relational. FMLP's G2 regime change further shows that this state must be measured after each update. Each round can therefore pair model improvement with state selection and, when possible, consolidation.

% \paragraph{Scope.}
% We study two generated updates for long-history users at fixed $\beta=0.5$ and two-model inference; G2 covers GRU4Rec and FMLP on three datasets. Reliable single-model consolidation and behavior over longer recursive chains remain open.

\section{Related Work}

\paragraph{Recursive self-improvement in recommendation.}
RSIR generates interactions and retrains successors under fidelity control~\cite{RSIR}; self-evolving systems revise objectives, code, and pipelines~\cite{SelfEvolveRec,YouTubeSelfEvolving}. We study the relational state created by such loops through a marginally matched contrast across their update boundary.

\paragraph{Consolidating relational state.}
RISE uses policy distillation for recursive improvement~\cite{RISE}. Recommender distillation transfers scores or ranks~\cite{RankingDistillation,HetComp}; EMKD combines sequential encoders~\cite{EMKD}, while Hydra and FLAME preserve ensemble diversity in one network~\cite{Hydra,FLAME}. These methods consolidate ensemble diversity; CGA diagnoses complementarity aligned with an update boundary under matched marginals.

\paragraph{Temporal and ensemble diversity.}
Continual recommendation balances stability and plasticity~\cite{Yoo2025ContinualRec}; ReLoop and meta-learning carry errors or experience forward~\cite{Cai2022ReLoop,Zhang2020SML}. Snapshot ensembles retain an optimization path~\cite{NeuSE}. CGA targets update-boundary complementarity through marginally matched cross- and within-generation pairs.

\section{Conclusion}

Recommendation recursive self-improvement produces learning states that need not collapse into the latest model. CGA exposes progress distributed across an update, while rank separation identifies the stronger pair family at selection time. Their variation across architectures and rounds, together with incomplete single-model consolidation, establishes relational state retention as a central design problem for self-improving recommender systems.

% \paragraph{Ethical considerations.}
% We use public interaction data. Deployed recursive loops should track popularity, minority-item exposure, and preference freshness.

% \clearpage
\bibliographystyle{ACM-Reference-Format}
\bibliography{references}

\end{document}